\newif\ifonecol
\onecolfalse

\documentclass[bibyear]{aa}

\usepackage{hyperref}

\usepackage[sort]{natbib}
\usepackage{txfonts}
\usepackage{graphicx}

\hypersetup{colorlinks,allcolors=blue}

\def\lum        {\ensuremath{L_\text{1.4~GHz}}}

\def\msolyr     {\ensuremath{\text{M}_{\odot}\,\text{yr}^{-1}}}

\def\whz        {\ensuremath{\text{W}\,\text{Hz}^{-1}}}
\def\mujybeam {\ensuremath{\mu\text{Jy}\,\text{beam}^{-1}}}
\def\mujy       {\ensuremath{\mu\text{Jy}}}
\def\vmax       {\ensuremath{V_\text{max}}}
\def\qtir       {\ensuremath{q_\text{TIR}}}

\authorrunning{Novak et al.}
\titlerunning{Submicrojansky radio source counts}

\usepackage{color}
\usepackage{amstext}

\begin{document}

\defcitealias{novak17}{N17}
\defcitealias{smolcic17c}{S17c}

\title{Constraints on submicrojansky radio number counts based on evolving VLA-COSMOS luminosity functions}

\author{
        M.~Novak\inst{1,2}, 
        V.~Smol\v{c}i\'{c}\inst{1},
        E.~Schinnerer\inst{2},
        G.~Zamorani\inst{3},
        I.~Delvecchio\inst{1},
        M.~Bondi\inst{4},
        J.~Delhaize\inst{1}
        }

\institute{
% 1
Department of Physics, Faculty of Science, University of Zagreb,  Bijeni\v{c}ka cesta 32, 10000 Zagreb, Croatia
\and
% 2
Max-Planck-Institut f\"{u}r Astronomie, K\"{o}nigstuhl 17, D-69117 Heidelberg, Germany
\and
% 3
INAF - Osservatorio Astronomico di Bologna, Via Piero Gobetti 93/3, I-40129 Bologna, Italy
\and
% 4
Istituto di Radioastronomia di Bologna - INAF, Via Piero Gobetti 101, 40129, Bologna, Italy
}

\date{Received ; accepted}

\abstract{
We present an investigation of radio luminosity functions (LFs) and number counts based on the Karl G. Jansky Very Large Array-COSMOS 3~GHz Large Project. The radio-selected sample of 7826 galaxies with robust optical/near-infrared counterparts with excellent photometric coverage allows us to construct the total radio LF since $z\sim5.7$. 
Using the Markov chain Monte Carlo algorithm, we fit the redshift dependent pure luminosity evolution model to the data and compare it with previously published VLA-COSMOS LFs obtained on individual populations of radio-selected star-forming galaxies and galaxies hosting active galactic nuclei classified on the basis of presence or absence of a radio excess with respect to the star-formation rates derived from the infrared emission. We find they are in excellent agreement, thus showing the reliability of the radio excess method in selecting these two galaxy populations at radio wavelengths. 
We study radio number counts down to submicrojansky levels drawn from different models of evolving LFs. We show that our evolving LFs are able to reproduce the observed radio sky brightness, even though we rely on extrapolations toward the faint end. Our results also imply that no new radio-emitting galaxy population is present below 1~\mujy. Our work suggests that selecting galaxies with radio flux densities between 0.1 and 10~\mujy\ will yield a star-forming galaxy in 90-95\% of the cases with a high percentage of these galaxies existing around a redshift of $z\sim2$, thus providing useful constraints for planned surveys with the Square Kilometer Array and its precursors.
}


\maketitle

\section{Introduction}
\label{sec:intro}

Radio emission in galaxies below rest-frame frequencies of 30~GHz is mostly synchrotron radiation arising from cosmic electrons gyrating in the galaxy magnetic fields, with a fraction of thermal emission due to free-free processes \citep[e.g.,][]{sadler89, condon92, clemens08,tabatabaei17}. 
Based on the source of acceleration of these cosmic electrons, galaxies can be separated into two categories, as follows.
%The acceleration source of these cosmic electrons separates galaxies into two categories, as follows.
The electrons are being accelerated by supernovae in star-forming (SF) galaxies, and by an accreting supermassive black hole (SMBH) in active galactic nuclei (AGN).
It is often not possible to make a clear distinction between these two types as many galaxies are composites.

Several classification schemes for radio AGN exist, but they are not fully consistent with each other as they are based either solely on properties observed in the radio or on a combination of host properties \citep[e.g.,][]{smolcic15conf}.
For example, one method uses the ratio of radio-to-optical flux densities to divide the AGN into radio-loud (RL) and radio-quiet (RQ) AGN \citep[e.g.,][]{bonzini13, padovani15}. A different method analyzes optical emission lines to classify AGN into low- and high-excitation galaxies \citep[LEGs and HEGs, respectively,  e.g.,][]{laing94,hardcastle06,smolcic09c}. We use the latter nomenclature for our radio sources to keep the AGN classes as general as possible, noting that the presence of a radio detection is not necessary for such a division since it is  based on the host properties \citep[for a review of the extensive nomenclature of radio AGN, see also ][]{padovani17}.
These observational constraints are used to probe the underlying physical mechanism, which can be best understood through SMBH accretion efficiency, that is, the amount of gravitational energy converted into radiation \citep[see][]{heckman14}. An AGN with a luminosity higher than 1 -- 10 \% of the Eddington limit can be considered radiatively efficient (this category largely overlaps with HEGs); it radiates energy mostly isotropically \citep[e.g.,][]{fanidakis11,heckman14}. On the other hand, radiatively inefficient AGN (overlapping with LEGs) have luminosities below 1 -- 10\% of the Eddington limit, and the energy is emitted mostly mechanically in the form of collimated jets \citep[see also][]{evans06,merloni08}. 
%radiatively inefficient AGN (overlapping with LEGs and RL AGN)
%radiatively efficient (this category largely overlaps with HEGs and RQ AGN)
The difference between the two types of AGN can also be seen in the properties of their galaxy hosts and their environments. Typically, HEGs have lower stellar and SMBH masses, but higher gas masses. Compared to LEGs, they also have bluer (green) colors and reside almost exclusively in less dense environments \citep[e.g.,][]{best12,gendre13}.
LEGs are usually hosted by massive red galaxies residing in denser environments \citep[see also][]{hale17}. 
These observations are consistent with the idea that SMBHs of HEGs are fueled by accretion of cold gas, which is possibly being enriched through galaxy mergers, while SMBHs of LEGs obtain their gas supply from the cluster halo \citep[for more details, see][]{best12}. It has also been shown that these two AGN types evolve differently with cosmic time. At lower luminosities, HEGs are less abundant, but evolve strongly (similar to SF galaxies), while LEGs show little or no evolution at all \citep[e.g.,][]{pracy16}. 

Analysis of emission lines in the optical spectra has proven to be a very successful method for identifying different types of AGN and separating them from SF galaxies, but the origin of the radio emission in AGN selected in this way might not be necessarily linked  to SMBH accretion processes or jets alone. 
Previous studies have shown that HEGs have higher dust temperatures, significant star formation rates, and obey the infrared (IR)-radio correlation valid for SF galaxies \citep[e.g.,][]{moric10,hardcastle13}. \cite{delvecchio17} find that only 30\% of AGN classified via X-ray or mid-IR signatures show a significant radio excess compared to the total IR emission arising from star formation. 
In the remainder of the sample that were not classified as X-ray/IR AGN, they showed that some AGN appear as "red and dead" galaxies in all bands except the radio \citep[see also][]{heckman14}.
When starting from a radio-selected sample, the division between the two AGN classes is less obvious because many galaxies are composites. The well-known radio loudness dichotomy emerges when radio emission is analyzed starting from an optically selected sample of AGN \citep[e.g.,][]{kimball11, balokovic12}.
For the purpose of using radio emission as a proxy for measuring star formation rates (SFRs) or AGN feedback, it is important to estimate which process dominates the radio emission: star formation processes, or SMBH accretion. 
%We do this by measuring the radio excess compared to the total star formation based IR emission
For this reason, we consider many HEGs as SF galaxies even though
they show a clear AGN contribution in  non-radio bands. Their radio emission, which is investigated here, is consistent with originating from star formation processes in the host galaxy (see also \citealt{smolcic17b} for a more detailed discussion of this).

One way to constrain how different populations add up to the total radio sky is through source number counts, that is, the number of galaxies contained in a solid angle of the sky with a given flux density \citep[for a review from a radio perspective, see, e.g.,][]{padovani16}.
First used as a tool for studying the geometry of the Universe, source number counts became a practical method for tracing galaxy evolution.
Such studies can also provide a statistical way to search for and describe new populations of galaxies \citep[e.g.,][]{longair66,franzen14}.
Flattening of Euclidean normalized number counts at sub-mJy flux densities at 1.4~GHz indicated the surfacing of a new galaxy population in the radio \citep[e.g.,][]{rowan-robinson93, seymour04, padovani09}, but the scatter in the observed counts from different datasets was significant \citep[e.g.,][]{hopkins03, heywood13}. There were also some indications that the 1.4~GHz counts could remain flat down to 10~\mujy\ or might even rise \citep[e.g.,][]{owen08}.
\cite{fixsen11} and \cite{vernstrom11} discussed in detail the implications of the observed radio excess from ARCADE\footnote{Absolute Radiometer for Cosmology, Astrophysics and Diffuse Emission} 2 sky brightness measurements, which suggested the existence of an abundant, but faint new radio population (possibly due to diffuse emission from clusters or halos or dark matter annihilation). However, recent deep VLA observations show evidence of a further steepening of the number counts below $\sim50~\mujy$ at 1.4~GHz
that is consistent with the idea of SF galaxies dominating the faint radio sky \citep[e.g.,][]{vernstrom16b, smolcic17a}.

There are several methods applicable for calculating source number counts. The traditional method relies on counting discrete sources in flux density bins (see, e.g., \citealt{dezotti10}). It is the most straightforward approach, although subject to biases, such as non-uniform $rms$ across the radio map, or resolution bias, and correct for such incompleteness issues \citep[e.g.,][]{hales14, vernstrom16a, smolcic17a}.
Another approach is a blind probability of deflection analysis $P(D)$  performed on a confusion-limited survey \citep[see][]{scheuer57}. This was recently used on deep Karl G. Jansky Very Large Array (VLA) data by \cite{condon12} and \cite{vernstrom14}. The method measures the bias in the noise and relies on a small number of assumptions.
The advantage of the $P(D)$ analysis is that it  can also model source counts well below the usual $5\sigma$ sensitivity limit.
However, because of its blind nature, it yields no information on specific sub-populations of galaxies. 
This method cannot be applied to high-resolution maps when the map is no longer dominated by confusion noise. In this case, one possible option is to stack the radio map at preset positions drawn from a catalog of priors derived through higher sensitivity (not necessarily radio) observations \citep[for a Bayesian approach, see, e.g.,][]{mitchell-wynne14, zwart15}. This method also allows probing the radio map below the nominal sensitivity threshold, but careful assessment of biases is required \citep[see also][]{zwart15b}.

If a cosmological model for the Universe is assumed, the radio number counts are a natural consequence of the underlying luminosity functions (LFs) \citep[e.g.,][]{condon89}. Recently, \cite{novak17}, hereafter \citetalias{novak17}, and \cite{smolcic17c}, hereafter \citetalias{smolcic17c}, published LFs of SF galaxies and AGN, respectively, using the VLA-COSMOS 3~GHz Large Project data \citep{smolcic17a}.
In this work we use LFs to estimate the number counts, relative fractions of different galaxy sub-populations, and redshift distributions of SF galaxies down to submicrojansky flux density levels.
Furthermore, we try to fit the total measured radio LF with different models of evolving LFs. Such a fit no longer relies on classifications of individual galaxies, but only on the analytical representation of the LF of the global population. This approach helps in mitigating biases introduced by misclassifications of galaxies.
The deep COSMOS field is optimized to probe faint distant sources that will predominantly be SF galaxies. 
To successfully distinguish the evolutions of HEGs and LEGs from the radio perspective, wide and shallow surveys are additionally needed to probe the bright luminosity end where the two AGN populations change their relative fractions ($\lum \gtrsim 10^{26}~\whz$, e.g., \citealt{pracy16}).
Constraints on the faint radio populations are useful for future radio surveys such as the SKA\footnote{Square Kilometer Array}, ASKAP\footnote{Australian SKA Pathfinder}, EMU\footnote{The Evolutionary Map of the Universe} and MeerKAT\footnote{Meer Karoo Array Telescope} \citep[see also][]{norris13}. 
%The usual paradigm is that multi-wavelength photometry is needed to classify galaxies as SF or AGN, however, in this paper we show that the future radio observations will provide a valuable and simple method to perform this classification based on the observed radio flux density alone.
In this paper we show that future radio observations will provide an additional valuable and simple  galaxy classification method based on the observed radio flux density alone.

In Section~\ref{sec:data} we briefly describe the radio and the ancillary data used throughout this work. In Section~\ref{sec:lumfun} we focus on methods for deriving the radio LF and its evolution through cosmic time. We also discuss biases that might affect our results. In Section~\ref{sec:counts} we discuss the radio number counts derived from radio LFs and implications for future radio surveys. Finally, in Section~\ref{sec:summary} we provide a brief summary.

We assume flat concordance lambda cold dark matter ($\Lambda$CDM) cosmology defined with a Hubble constant of  $H_0=70$\,km\,s$^{-1}$\,Mpc$^{-1}$, dark energy density of $\Omega_\Lambda=0.7$, and matter density of $\Omega_\text{m}=0.3$. For the radio spectral energy distribution we assume a simple power law described as $S_\nu\propto\nu^\alpha$, where $S_\nu$ is the flux density at frequency $\nu$ and $\alpha$ is the spectral index. If not explicitly stated otherwise, $\alpha=-0.7$ is assumed.

%\cite{whittam17}
%\cite{mcalpine13}

%IR-RADIO correlation?

\section{Data}
\label{sec:data}

The data used throughout the paper are radio selected and have been cross-correlated with the multiwavelength optical/near-infrared (NIR) and value-added catalogs. 
The radio data were taken with the VLA at 3~GHz, reaching a sensitivity of 2.3~\mujybeam\ at a resolution of $0\farcs75$. The observational setup, data reduction, source extraction, and completeness simulations are described in detail in \cite{smolcic17a}.
In short, 384 hours of observations were conducted with the VLA in $S$-band (2 -- 4~GHz continuum) to uniformly cover the 2~deg$^2$ COSMOS field with 192 pointings. Both the VLA A and C configurations were used. The scatter of the flux calibration in the different pointings is 5\% around the mean, ensuring good flux measurements. The data were imaged with the multiscale multifrequency synthesis \citep{rau11} to obtain the intensity (Stokes I) radio map. The radio components above $5\sigma$ were extracted using \textsc{blobcat} \citep{hales12}, and were further inspected by eye where necessary. The final source catalog contains around 5\,000 galaxies per square degree. The completeness of the 3~GHz catalog was estimated through Monte Carlo simulations of mock sources, and it was given as a function of flux density. The catalog is 60\% complete below 20~\mujy, 95\% above 40~\mujy,\ and a 100\% completeness is assumed above 100~\mujy.

Radio sources were assigned counterparts \citep[for details see][]{smolcic17b} primarily from the COSMOS2015 catalog \citep{laigle16}.
The COSMOS2015 catalog contains the high-quality multiwavelength photometry of $\sim$800\,000 sources across more than 30 bands
from near-ultraviolet (NUV) to NIR obtained from GALEX, UltraVISTA DR2, Subaru/Hyper-Suprime-Cam, and the SPLASH Spitzer legacy program. Sources were identified in the $\chi^2$ weighted stack of $z^{++} Y J H K_s$ bands. The photometry extraction was performed with \textsc{SExtractor} within 2$\arcsec$ and 3$\arcsec$ apertures on individual bands that were previously homogenized to a common point-spread-function (resolution of 0$\farcs$8).
For higher counterpart completeness, additional sources were considered from the $i$-band selected catalog \citep{capak07} and the 3.6~$\mu$m \emph{Spitzer}/IRAC\footnote{Infrared Array Camera} catalog \citep{saunders90}.

The process of the cross-correlation of the 3~GHz radio and multiwavelength sources is explained in detail in \cite{smolcic17b}. Associations of radio sources to NIR/optical sources were based on the nearest-neighbor matching within a search radius of 0$\farcs$8 for the COSMOS2015/$i$-band, and a radius of 1$\farcs$7 for the IRAC catalog. Additionally, false-match probabilities were drawn from Monte Carlo simulations using a background model that mimics the observed $m_{3.6~\mu\text{m}}$ magnitude distribution of radio counterparts. 
%This approach takes into account the effect of blocking, i.e. fainter NIR/optical sources not being identified in the presence of a nearby bright radio counterpart.
The entire process resulted in 8\,035 radio sources  with optical/NIR counterparts across an area of 1.77~deg$^2$ free from saturation and bright stars' contamination (7\,729 from COSMOS2015, 97 from $i$-band and 209 from the IRAC catalogs).
We do not use sources assigned an IRAC counterpart as robust photometric redshifts cannot be obtained from the IRAC data alone.

%Third, a false match probability was assigned to every counterpart match based on Monte Carlo simulations. For the simulations we matched sources from mock catalogs that realistically reflect the mag- nitude, and separation distributions expected for the counter- parts of our radio sources (see Appendix A.2 for details). These steps led to three counterpart candidate catalogs, each, respec- tively, containing the 3 GHz sources matched to sources from the 1) COSMOS2015 (see Appendix A.3), 2) i-band (see Ap- pendix A.4), 3) IRAC catalogs (see Appendix A.5), and with a false match probability computed for each match.

% 7729 c15; 97 iband; 209 irac
% 9986 vla3 in cosmos
% 8696 in gud area

A reliable photometric redshift estimate is available for 7\,826 radio sources associated with counterparts in the COSMOS2015 or the $i$-band selected catalog. These sources form our final radio selected sample used throughout. 
This final sample represents 90\% of all cataloged radio sources. Of the remaining 10\%, one half of the sources have a low signal-to-noise ratio (S/N) of $<$ 6, making many of them likely candidates for spurious detections.
It was estimated by \cite{smolcic17a} that 3\% of the sources in the 3~GHz radio catalog are expected to be spurious (mostly those with S/N $<$ 5.5).
By taking into account the false-detection probability of the VLA-COSMOS 3~GHz radio survey, we estimate that 7\% of real radio sources were not assigned a counterpart (this includes IRAC associations), introducing a small negative bias into our counting statistics. It is likely that the redshift distribution of these unaccounted sources is not uniform across all redshifts, but skewed toward high redshift. This would imply somewhat higher densities of sources at high redshifts at a given luminosity than suggested by the data. 
In the absence of an actual redshift distribution of these sources, we do not attempt to correct for this bias.

Spectroscopic redshift is available for 35\% of our sources, mostly below $z<1.5$.
The median accuracy of the photometric redshifts in our sample is $\Delta z/(1+z_\text{spec})=0.01 (0.04)$, with a catastrophic failure rate of 4\% (12\%) for redshift range $z>0$ ($z>1.5$).

\section{Luminosity functions and evolution}
\label{sec:lumfun}

We first describe the methods used to measure the total radio LFs from the VLA-COSMOS 3~GHz data. Then, we briefly summarize the results from \citetalias{novak17} and \citetalias{smolcic17c}, where the authors focused on individual galaxy populations: SF and AGN, respectively. Finally, we fit different evolution models of the total radio LF and discuss the various models as well as potential biases in the analysis.

\subsection{Measuring the total radio luminosity function}
\label{sec:lumfun_total}

\begin{figure*}
\centering
\includegraphics[width=\linewidth]{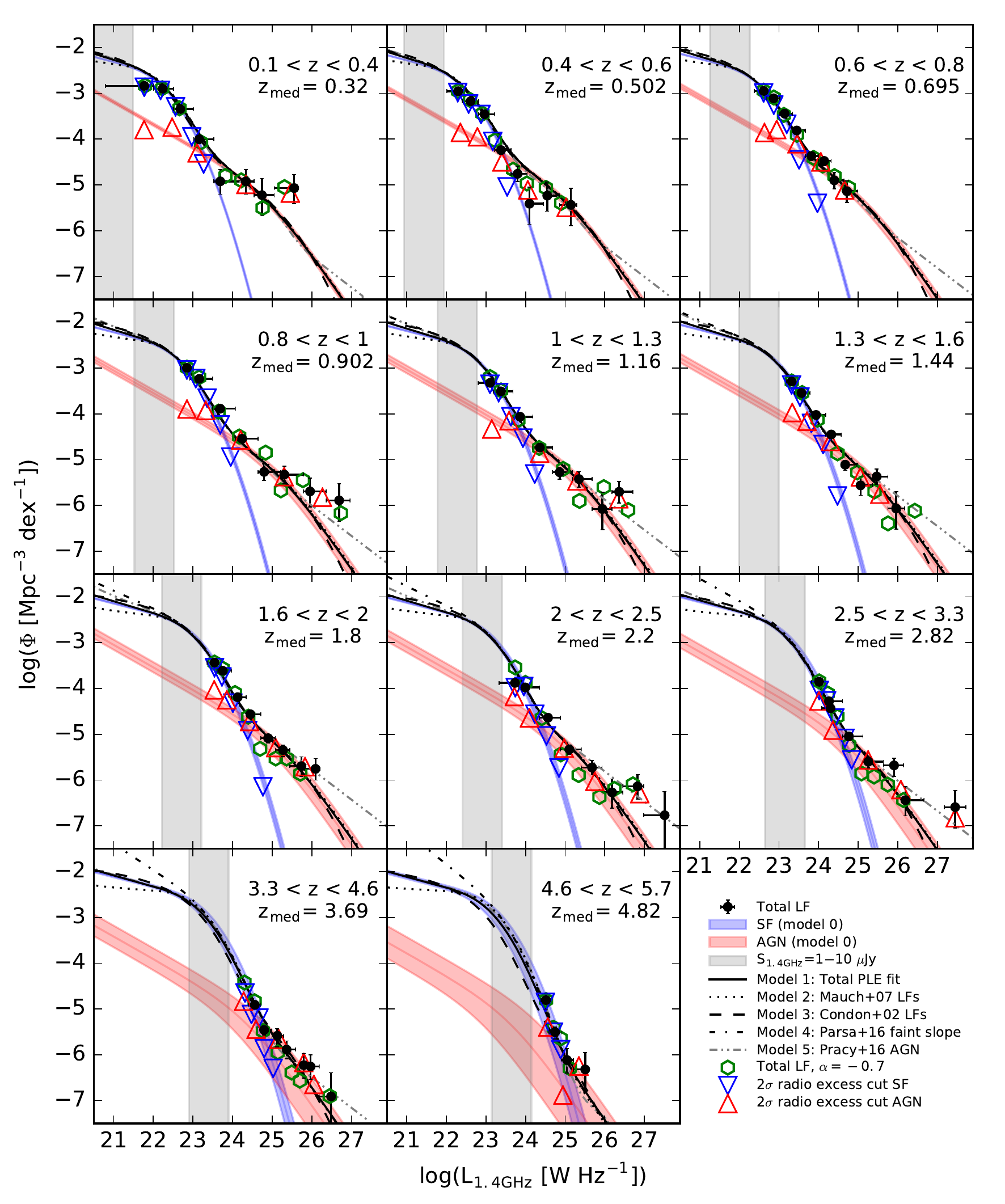}
\caption{Total radio luminosity functions at different cosmic epochs. Black points show LFs derived using the \vmax\ method  (see Section~\ref{sec:lumfun_total}). The blue and red shaded areas show the $\pm 3\sigma$ ranges of the best-fit evolution for the individual SF and AGN populations, respectively (outlined in Section~\ref{sec:lumfun_pop}). The black line is the fit to the total radio LF, as explained in Section~\ref{sec:fit_total}. Other models used are indicated in the legend. In all of these models we assume the local ($z\approx0$) LFs  from the respective papers and fit for the PLE parameters as in Eq.~\ref{eq:lf_total}. The only exception is model 4, where the shape of the faint slope of the SF LF changes with redshift.
The vertical gray shaded area shows the luminosity decade that contributes to the radio source counts between 1 and 10~\mujy.}
\label{fig:lfgrid}
\end{figure*}

To calculate the total radio LFs, we followed the procedure that
was used in \citetalias{novak17}. The method is based on computing the maximum observable volume, \vmax, for each source \citep{schmidt68}
and simultaneously applying completeness corrections that take into account the nonuniform $rms$ noise and the resolution bias \citepalias[see Section~3.1 in][]{novak17}. We did not apply any statistical corrections for radio sources without assigned optical counterparts as the redshift distribution is unknown, but we expect these effects to be less than 10\% (see also Section~\ref{sec:data}). 
In order to obtain the rest-frame luminosity of a radio source, a spectral index must be assumed. 
The spectral index of a source is calculated between 1.4 and 3~GHz, if a 1.4~GHz detection \citep{schinnerer10} is available, and assumed to remain unchanged at all radio frequencies. Such spectral indices are available for a quarter of the sample. For sources detected only at 3~GHz, we assumed $\alpha=-0.7$, which corresponds to the average spectral index of the entire 3~GHz population \citep[see Sect.~4 in][]{smolcic17a}.\
% (see Section~6.4 in \citetalias{novak17} for additional discussion regarding biases of this choice).
Redshift bins were chosen to be large enough not to be affected severely by photometric redshift uncertainty.
Luminosity bins in each redshift bin span exactly the observed luminosity range of the data.
The lowest luminosity bin spans from the faintest observed source to the $5\sigma$ detection threshold  at the upper redshift limit (corresponding to $5\times 2.3~\mujybeam$ at 3~GHz) so that possible issues due to poorer sampling can be mitigated more easily.
The reported luminosity for each LF is the median luminosity of the sources within the bin. The horizontal bars in the plots correspond to the width of the bin. The total radio LFs are shown in Fig.~\ref{fig:lfgrid} (black points) and are also listed in Table~\ref{tab:lumfun_vmax}.
In an attempt to quantify the bias introduced by an imperfect radio K correction, which will mostly affect the bright end of the LF, the total radio LF was also calculated assuming a spectral index of $\alpha=-0.7$ for all sources (see green hexagons in Fig.~\ref{fig:lfgrid}). We discuss this further below.

\begin{table*}
\begin{center}
\caption{Luminosity functions of the total radio-selected sample obtained with the \vmax\ method.}
% for publication
%\small
\renewcommand{\arraystretch}{1.5}
\ifonecol \scriptsize \fi
\begin{tabular}[t]{c c c}
\hline
$\log\lum$ & $\log\Phi$& N \\
$[\text{W}\,\text{Hz}^{-1}]$ & $[\text{Mpc}^{-3}\,\text{dex}^{-1}]$ & \\
\hline
$z=0.1-0.4$ & & \\
\hline
21.77$_{-0.99}^{+0.23}$ & -2.84$_{-0.068}^{+0.080}$ & 202 \\
22.24$_{-0.24}^{+0.27}$ & -2.90$_{-0.023}^{+0.024}$ & 352 \\
22.68$_{-0.17}^{+0.34}$ & -3.34$_{-0.035}^{+0.038}$ & 145 \\
23.16$_{-0.14}^{+0.37}$ & -4.00$_{-0.070}^{+0.083}$ & 33 \\
23.69$_{-0.16}^{+0.35}$ & -4.92$_{-0.25}^{+0.28}$ & 4 \\
24.34$_{-0.30}^{+0.21}$ & -4.92$_{-0.25}^{+0.28}$ & 4 \\
24.74$_{-0.20}^{+0.31}$ & -5.22$_{-0.37}^{+0.45}$ & 2 \\
25.56$_{-0.50}^{+0.030}$ & -5.07$_{-0.30}^{+0.34}$ & 3 \\
\hline
$z=0.4-0.6$ & &\\
\hline
22.30$_{-0.29}^{+0.11}$ & -2.95$_{-0.043}^{+0.048}$ & 151 \\
22.61$_{-0.21}^{+0.20}$ & -3.17$_{-0.026}^{+0.027}$ & 271 \\
22.96$_{-0.15}^{+0.26}$ & -3.46$_{-0.033}^{+0.036}$ & 165 \\
23.38$_{-0.16}^{+0.25}$ & -4.24$_{-0.074}^{+0.089}$ & 29 \\
23.80$_{-0.17}^{+0.24}$ & -4.76$_{-0.16}^{+0.17}$ & 9 \\
24.10$_{-0.060}^{+0.35}$ & -5.41$_{-0.37}^{+0.45}$ & 2 \\
24.55$_{-0.099}^{+0.31}$ & -5.23$_{-0.30}^{+0.34}$ & 3 \\
25.14$_{-0.29}^{+0.15}$ & -5.44$_{-0.37}^{+0.45}$ & 2 \\
\hline
$z=0.6-0.8$ & & \\
\hline
22.61$_{-0.24}^{+0.080}$ & -2.95$_{-0.054}^{+0.062}$ & 190 \\
22.86$_{-0.17}^{+0.15}$ & -3.11$_{-0.023}^{+0.024}$ & 354 \\
23.14$_{-0.12}^{+0.20}$ & -3.44$_{-0.030}^{+0.032}$ & 205 \\
23.45$_{-0.10}^{+0.22}$ & -3.81$_{-0.043}^{+0.048}$ & 92 \\
23.82$_{-0.15}^{+0.17}$ & -4.37$_{-0.079}^{+0.096}$ & 26 \\
24.14$_{-0.15}^{+0.18}$ & -4.48$_{-0.086}^{+0.11}$ & 21 \\
24.40$_{-0.080}^{+0.24}$ & -4.90$_{-0.17}^{+0.18}$ & 8 \\
24.71$_{-0.074}^{+0.28}$ & -5.14$_{-0.22}^{+0.25}$ & 5 \\
\hline
$z=0.8-1.0$ & &\\
\hline
22.85$_{-0.16}^{+0.072}$ & -2.99$_{-0.038}^{+0.042}$ & 184 \\
23.16$_{-0.24}^{+0.33}$ & -3.24$_{-0.017}^{+0.018}$ & 647 \\
23.69$_{-0.19}^{+0.38}$ & -3.89$_{-0.032}^{+0.035}$ & 177 \\
24.24$_{-0.17}^{+0.40}$ & -4.54$_{-0.062}^{+0.073}$ & 42 \\
24.80$_{-0.15}^{+0.42}$ & -5.27$_{-0.17}^{+0.18}$ & 8 \\
25.31$_{-0.090}^{+0.48}$ & -5.33$_{-0.19}^{+0.20}$ & 7 \\
25.96$_{-0.16}^{+0.41}$ & -5.69$_{-0.30}^{+0.34}$ & 3 \\
26.69$_{-0.33}^{+0.28}$ & -5.89$_{-0.37}^{+0.45}$ & 2 \\

\hline
\end{tabular}
\renewcommand{\arraystretch}{1.5}
\begin{tabular}[t]{c c c}
\hline
$\log\lum$ & $\log\Phi$ & N \\
$[\text{W}\,\text{Hz}^{-1}]$ & $[\text{Mpc}^{-3}\,\text{dex}^{-1}]$ & \\  
\hline
$z=1.0-1.3$ & & \\
\hline
23.10$_{-0.30}^{+0.080}$ & -3.32$_{-0.041}^{+0.045}$ & 228 \\
23.38$_{-0.19}^{+0.30}$ & -3.51$_{-0.019}^{+0.019}$ & 542 \\
23.86$_{-0.18}^{+0.32}$ & -4.06$_{-0.031}^{+0.033}$ & 190 \\
24.36$_{-0.18}^{+0.32}$ & -4.74$_{-0.062}^{+0.072}$ & 43 \\
24.86$_{-0.18}^{+0.32}$ & -5.28$_{-0.11}^{+0.15}$ & 12 \\
25.35$_{-0.17}^{+0.33}$ & -5.43$_{-0.16}^{+0.17}$ & 9 \\
25.94$_{-0.26}^{+0.24}$ & -6.08$_{-0.37}^{+0.45}$ & 2 \\
26.36$_{-0.18}^{+0.34}$ & -5.70$_{-0.22}^{+0.25}$ & 5 \\
\hline
$z=1.3-1.6$ & & \\
\hline
23.32$_{-0.14}^{+0.068}$ & -3.29$_{-0.037}^{+0.040}$ & 173 \\
23.57$_{-0.18}^{+0.22}$ & -3.54$_{-0.021}^{+0.022}$ & 432 \\
23.94$_{-0.15}^{+0.24}$ & -4.02$_{-0.031}^{+0.033}$ & 188 \\
24.32$_{-0.14}^{+0.25}$ & -4.45$_{-0.049}^{+0.055}$ & 72 \\
24.67$_{-0.10}^{+0.29}$ & -5.11$_{-0.098}^{+0.13}$ & 16 \\
25.06$_{-0.095}^{+0.30}$ & -5.56$_{-0.20}^{+0.22}$ & 6 \\
25.47$_{-0.11}^{+0.28}$ & -5.37$_{-0.16}^{+0.17}$ & 9 \\
25.96$_{-0.21}^{+0.21}$ & -6.07$_{-0.37}^{+0.45}$ & 2 \\
\hline
$z=1.6-2.0$ & & \\
\hline
23.54$_{-0.19}^{+0.067}$ & -3.43$_{-0.054}^{+0.061}$ & 192 \\
23.75$_{-0.14}^{+0.22}$ & -3.61$_{-0.019}^{+0.020}$ & 494 \\
24.13$_{-0.15}^{+0.21}$ & -4.19$_{-0.032}^{+0.035}$ & 173 \\
24.45$_{-0.11}^{+0.26}$ & -4.56$_{-0.049}^{+0.055}$ & 73 \\
24.90$_{-0.19}^{+0.18}$ & -5.09$_{-0.081}^{+0.099}$ & 24 \\
25.27$_{-0.19}^{+0.17}$ & -5.34$_{-0.11}^{+0.14}$ & 13 \\
25.74$_{-0.29}^{+0.075}$ & -5.70$_{-0.20}^{+0.22}$ & 6 \\
26.10$_{-0.28}^{+0.11}$ & -5.75$_{-0.22}^{+0.25}$ & 5 \\
\hline
$z=2.0-2.5$ & & \\
\hline
23.73$_{-0.40}^{+0.093}$ & -3.88$_{-0.040}^{+0.044}$ & 168 \\
23.99$_{-0.17}^{+0.36}$ & -3.98$_{-0.021}^{+0.022}$ & 410 \\
24.57$_{-0.21}^{+0.31}$ & -4.64$_{-0.040}^{+0.044}$ & 112 \\
25.10$_{-0.23}^{+0.30}$ & -5.33$_{-0.085}^{+0.11}$ & 23 \\
25.68$_{-0.27}^{+0.25}$ & -5.73$_{-0.15}^{+0.16}$ & 10 \\
26.18$_{-0.25}^{+0.28}$ & -6.27$_{-0.30}^{+0.34}$ & 3 \\
26.83$_{-0.37}^{+0.16}$ & -6.14$_{-0.25}^{+0.28}$ & 4 \\
27.51$_{-0.53}^{+0.028}$ & -6.77$_{-0.52}^{+0.76}$ & 1 \\

\hline
\end{tabular}
\renewcommand{\arraystretch}{1.5}
\begin{tabular}[t]{c c c}
\hline
 $\log\lum$ & $\log\Phi$ & N \\
   $[\text{W}\,\text{Hz}^{-1}]$ & $[\text{Mpc}^{-3}\,\text{dex}^{-1}]$ & \\ 
 \hline
$z=2.5-3.3$ & & \\
\hline
24.01$_{-0.20}^{+0.076}$ & -3.85$_{-0.043}^{+0.048}$ & 172 \\
24.26$_{-0.17}^{+0.34}$ & -4.28$_{-0.024}^{+0.025}$ & 327 \\
24.76$_{-0.16}^{+0.36}$ & -5.05$_{-0.051}^{+0.058}$ & 67 \\
25.26$_{-0.14}^{+0.38}$ & -5.60$_{-0.10}^{+0.13}$ & 17 \\
25.91$_{-0.27}^{+0.24}$ & -5.68$_{-0.15}^{+0.24}$ & 12 \\
26.19$_{-0.042}^{+0.47}$ & -6.44$_{-0.30}^{+0.34}$ & 3 \\
27.45$_{-0.27}^{+0.27}$ & -6.59$_{-0.37}^{+0.45}$ & 2 \\
\hline
$z=3.3-4.6$ & & \\
\hline
24.30$_{-0.22}^{+0.097}$ & -4.43$_{-0.070}^{+0.084}$ & 75 \\
24.56$_{-0.16}^{+0.14}$ & -4.91$_{-0.051}^{+0.058}$ & 66 \\
24.80$_{-0.10}^{+0.20}$ & -5.46$_{-0.087}^{+0.11}$ & 21 \\
25.13$_{-0.14}^{+0.16}$ & -5.58$_{-0.16}^{+0.24}$ & 11 \\
25.37$_{-0.072}^{+0.23}$ & -5.89$_{-0.19}^{+0.20}$ & 7 \\
25.80$_{-0.21}^{+0.092}$ & -6.23$_{-0.25}^{+0.28}$ & 4 \\
25.97$_{-0.080}^{+0.22}$ & -6.26$_{-0.25}^{+0.28}$ & 4 \\
26.49$_{-0.30}^{+0.026}$ & -6.91$_{-0.52}^{+0.76}$ & 1 \\
\hline
$z=4.6-5.7$ & & \\
\hline
24.52$_{-0.13}^{+0.077}$ & -4.81$_{-0.14}^{+0.20}$ & 14 \\
24.75$_{-0.15}^{+0.16}$ & -5.51$_{-0.11}^{+0.15}$ & 12 \\
25.05$_{-0.15}^{+0.16}$ & -6.12$_{-0.25}^{+0.28}$ & 4 \\
25.51$_{-0.30}^{+0.035}$ & -6.32$_{-0.37}^{+0.45}$ & 2 \\

\hline
\end{tabular}
\label{tab:lumfun_vmax}
\end{center}
\end{table*}

\subsection{Star-forming galaxies and AGN}
\label{sec:lumfun_pop}

\citetalias{novak17} and \citetalias{smolcic17c} classified galaxies based on the radio emission excess compared to the IR based SFR, following the prescription presented in \cite{delvecchio17}. This method was aimed at distributing galaxies into two categories depending on the physical process likely producing the radio emission. Galaxies that exhibit a $3\sigma$ significant radio excess, namely
\begin{equation}
\log\left(\frac{\lum[\whz]}{\text{SFR}_\text{IR}[\msolyr]}\right)>a(1+z)^b,
\label{eq:excess_cut}
\end{equation}
where $a=22.0$ and $b=0.013$, owe at least 80\% of their radio emission to the AGN, while the rest is due to star formation. The percentage of AGN contribution to the total radio emission corresponds to the offset from the peak of the $\log(\lum/\text{SFR}_\text{IR})$ distribution at a given redshift.
In the above relation, \lum\ is the radio luminosity and SFR$_\text{IR}$ is the star formation rate based on the integrated 8 -- 1\,000~$\mu$m SED, with any AGN component subtracted, using the \cite{kennicutt98} relation and the \cite{chabrier03} initial mass function.
Galaxies below such a defined radio excess threshold might still host an AGN (visible in X-rays for example), but the radio emission originates mostly in star formation processes. 
The peak of the \lum/SFR$_\text{IR}$ distribution shifts to higher values with increasing redshift, hence the need for a redshift-dependent description of the radio excess. Further evidence for the necessity of this redshift evolution arises from the analysis of the IR-radio correlation performed by \cite{delhaize17}, where the  authors find a decreasing trend in the IR-radio correlation parameter \qtir\ with redshift \citep[see also][]{ivison10,magnelli15}.

%We select 1846 galaxies satisfying Eq.~\ref{eq:excess_cut} as AGN, and 5980 galaxies below the same threshold as SF galaxies.

Radio LFs across cosmic time are usually described by a local LF evolved in luminosity, or density, or both \citep[e.g.,][]{condon84}. Following \citetalias{novak17} we can parametrize the redshift-dependent evolution with two parameters for density evolution ($\alpha_D, \beta_D$) and two parameters for luminosity evolution ($\alpha_L, \beta_L$) to obtain
\begin{equation}
\Phi(L,z, \alpha_L, \beta_L, \alpha_D, \beta_D) =  (1+z)^{\alpha_D+z\cdot\beta_D}\times\Phi_{0} \left[ \frac{L}{(1+z)^{\alpha_L+z\cdot\beta_L}}
\right],
\label{eq:lfevol_model}
\end{equation}
where $\Phi_{0}$ is the local LF. The linear form $\alpha+z\beta$ of the evolution parameter was chosen for its simplicity and its good fit to the data. Alternative multi-parameter descriptions also exist in the literature \citep[see, e.g.,][]{padovani15,gruppioni13}.
The shape and the evolution of the LF depend on the galaxy population type. \citetalias{novak17} used a power-law plus lognormal shape of the local LF for SF galaxies obtained as the best fit of the combined data from \cite{condon02}, \cite{best05} and \cite{mauch07}
\begin{equation}
\Phi_0^{\text{SF}}(L)=\Phi_\star\left(\frac{L}{L_\star}\right)^{1-\alpha} \exp\left[-\frac{1}{2\sigma^2}\log^2\left(1+\frac{L}{L_\star}\right)\right],
\label{eq:lflocal_sf}
\end{equation}
where $\Phi_\star=3.55\times10^{-3}~\text{Mpc}^{-3}\text{dex}^{-1}$, $L_\star=1.85\times10^{21}~\whz$, $\alpha=1.22$, and $\sigma=0.63$. 
They report the best-fit pure luminosity evolution (PLE) parameters $\alpha_L^{\text{SF}}=3.16$ and $\beta_L^{\text{SF}}=-0.32$, valid for $z<5.7$.

\citetalias{smolcic17c} used the double power-law shape reported in \cite{mauch07} to describe the local AGN LF,
\begin{equation}
\Phi_0^{\text{AGN}}(L) = \dfrac{\Phi_\star}{(L_\star/L)^\alpha + (L_\star/L)^\beta},
\label{eq:lflocal_agn}
\end{equation}
where $\Phi_\star = \frac{1}{0.4}10^{-5.5}$~Mpc$^{-3}$~dex$^{-1}$, $L_\star=10^{24.59}$~W~Hz$^{-1}$, $\alpha =-1.27$, and $\beta=-0.49$.  They report the best-fit PLE parameters 
%$\alpha_L^{\text{AGN}}=2.88\pm0.82$ and  $\beta_L^{\text{AGN}}=-0.84\pm0.34$,
$\alpha_L^{\text{AGN}}=2.88$ and  $\beta_L^{\text{AGN}}=-0.84$,
valid for $z<5.5$. They also calculated the best-fit pure density evolution (PDE), which is consistent with their PLE. In Fig.~\ref{fig:lfgrid} we show the luminosity-evolved LFs (using Eq.~\ref{eq:lfevol_model} with $\alpha_D=0$ and $\beta_D=0$) of the SF galaxies and AGN (Equations~\ref{eq:lflocal_sf} and \ref{eq:lflocal_agn}) as blue and red areas, respectively.

\subsection{Fitting the total luminosity function}
\label{sec:fit_total}

Following \cite{mcalpine13}, we fit the total radio LF with a combination of different LFs. 
Both PLE and PDE models are common in the literature \citep[e.g.,][]{condon84, sadler02, gruppioni13}, but the true evolution might be a combination of both of these extremes \citep[see, e.g.,][]{yuan16}, with a possible luminosity-dependent evolution as well \citep[see, e.g.,][]{fotopoulou16}.

It was noted by \citetalias{novak17} that the PDE of SF galaxies would push the densities to very high numbers, thus making them inconsistent with the observed cosmic star formation rate densities. This is a consequence of the fact that our data can constrain only the bright log-normal part of the SF LF. For AGN it was shown by \citetalias{smolcic17c} that the PDE and PLE models are similar, mostly because the shape of the LF does not deviate strongly from a simple power law at the observed luminosities. Considering the above reasoning while also trying to keep the parameter space degeneracy to a minimum, we decided to use only the PLE for our analysis. 

We constructed a four--parameter redshift-dependent pure luminosity evolution model with two parameters for the SF and AGN populations. The total LF is defined by the following evolutionary model:
\begin{equation}
\begin{split}
\Phi(&L,z,\alpha_L^{\text{SF}}, \beta_L^{\text{SF}}, \alpha_L^{\text{AGN}}, \beta_L^{\text{AGN}}) =\\ &=\quad\Phi_{0}^{\text{SF}} \left[ \frac{L}{(1+z)^{\alpha_L^{\text{SF}}+z\cdot\beta_L^{\text{SF}}}}\right]+
\Phi_{0}^{\text{AGN}} \left[ \frac{L}{(1+z)^{\alpha_L^{\text{AGN}}+z\cdot\beta_L^{\text{AGN}}}}\right],
\label{eq:lf_total}
\end{split}
\end{equation}
where $\Phi_{0}^{\text{SF}}$ and $\Phi_{0}^{\text{AGN}}$ are the local LFs for the two galaxy populations.

\begin{figure}
\centering
\includegraphics[width=\columnwidth]{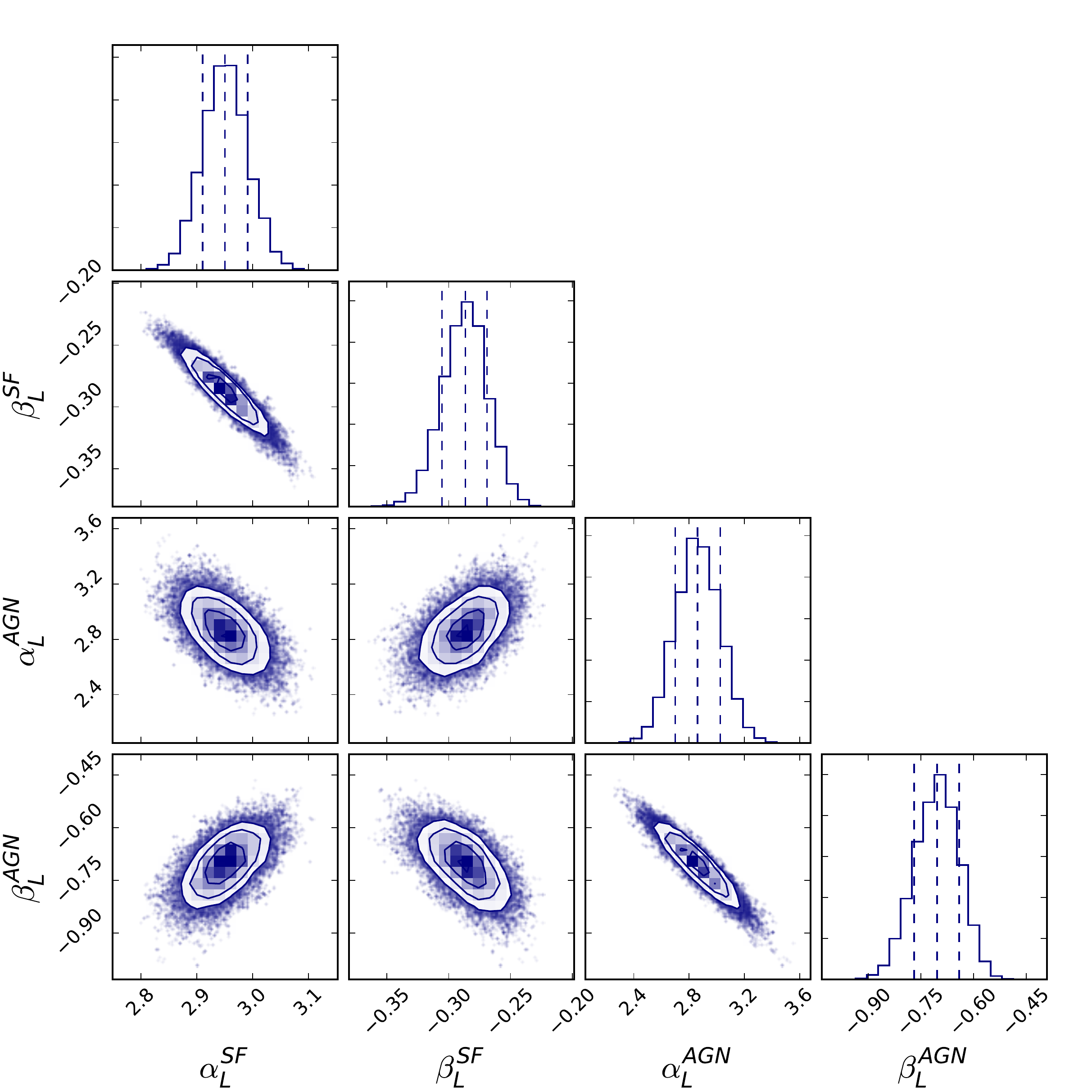}
\caption{Covariance maps of four fitting parameters used to describe the total radio LF evolution. Probability distribution histograms are also shown for each parameter with 16th, 50th and 84th percentiles marked by vertical dashed lines.}
\label{fig:corner}
\end{figure}

We used the Markov chain Monte Carlo (MCMC) algorithm, available in the Python package \textsc{emcee} \citep{foreman13}, to perform a multi-variate fit to the data in Table~\ref{tab:lumfun_vmax}. 
%The evolutionary model is described with Eq.~\ref{eq:lf_total} and the local LFs are defined in Eqs.~\ref{eq:lflocal_sf} and \ref{eq:lflocal_agn}. 
Covariance maps of the four fitted parameters are shown in Fig.~\ref{fig:corner}.
This plot shows that the total fit shows a slight degeneracy between SF and AGN populations, but the intersection of AGN and SF luminosity functions is well constrained by our data (see Fig.~\ref{fig:lfgrid}), and we therefore consider our fit robust despite this small degeneracy.
For an individual population, a higher $\alpha$ parameter correlates strongly with a steeper (more negative) value of $\beta$. The redshift dependence of the total evolution parameter $(\alpha+z\cdot\beta)$ is necessary to describe the observations at all redshifts. 
The best-fit parameter values are listed in Table~\ref{tab:evol} as model 1, which is also shown in Fig.~\ref{fig:lfgrid} (black line).

Since the shape of the LF is defined analytically, it extends beyond the observed data. The extrapolation toward the faint end is particularly important (i.e., below \lum<10$^{23}~\whz$ at $z>1$) because the densities of galaxies are the highest in this range. A change in the faint-end slope of the LF can easily change the relative fractions of galaxy populations at faint flux densities. The only way to mitigate this extrapolation issue is to obtain deeper data. However, in the absence of such observations, we made use of different LFs. We performed additional MCMC fits using analytical SF and AGN LFs from \cite{condon02}, where both LFs are hyperbolic in shape, \cite{mauch07}, and the redshift-dependent faint power-law slope for SF galaxies derived by \cite{parsa16}. The best-fit evolution parameters are listed in Table~\ref{tab:evol}, and the models are also shown in Fig.~\ref{fig:lfgrid}. 
%The effects these LF choices have on the final radio number counts are discussed in a later section.

We additionally tried decomposing the AGN population into its two subpopulations (HEGs and LEGs) by adding a second AGN LF to Eq.~\ref{eq:lf_total}. 
In this model we use the local LFs ($0.005< z< 0.3$) of LEGs and HEGs obtained by \cite{pracy16}, which are based on the Large Area Radio Galaxy Evolution Spectroscopic Survey (LARGESS), and fit for the evolutionary parameters as before.
We performed a six-parameter MCMC fit (two PLE parameters for each subpopulation) in an attempt to constrain the evolution of the total radio LF $\Phi(L,z,\alpha_L^{\text{SF}}, \beta_L^{\text{SF}}, \alpha_L^{\text{HEG}}, \beta_L^{\text{HEG}}, \alpha_L^{\text{LEG}},
\text{and }\beta_L^{\text{LEG}})$.
%We performed a six-parameter MCMC fit (two PLE parameters for each subpopulation) in an attempt to better understand the high luminosity end of our radio LFs. 
The result is shown in Fig.~\ref{fig:lfgrid} and is discussed in the next section.

\subsection{Discussing the fitted models and biases}
\label{sec:lf_biases}

One advantage of performing a fit to the total radio LF is that the results are no longer sensitive to the galaxy classification method. However, it is important to assume a proper shape of the LFs for distinct galaxy populations in order to obtain a meaningful evolution fit. For this reason, the shape of the LF for a given population is fixed to the local LF, which is usually constrained across at least five orders of magnitudes in luminosities. The  best-fit evolution parameters obtained from the MCMC on the total radio LF are consistent within the $3\sigma$ errors with $\chi^2$ fits performed on individual populations (see shaded areas in Fig.~\ref{fig:lfgrid}). 
This good agreement between the results of different analyses implies that regardless of the specifics of the galaxy classification methods, if the local LFs for SF and AGN galaxies are well described with Eq.~\ref{eq:lflocal_sf} and \ref{eq:lflocal_agn}, our PLE parameters are robust.
However, the total radio LFs suggest slightly higher AGN densities than those obtained in \citetalias{smolcic17c}. Some redshift bins have some high-luminosity outliers, that is, data points that are above the best-fit LF. We now discuss potential biases that might cause these discrepancies.

\subsubsection{Spectral index bias}

The largest uncertainty in measuring the radio LF, especially evident at higher redshifts, is the imperfect knowledge of the radio SED. Under the assumption of a simple power law, an offset in spectral index of $\Delta\alpha=0.1$ would change the 1.4~GHz rest-frame luminosity by 0.08 and 0.11~dex for $z=2$ and $z=5$, respectively.
Previous studies have shown that there is a $\sim0.4$ symmetric spread of spectral indices around the mean value \citep[e.g.,][]{kimball08}. 
A single power-law approximation might not hold if the radio SED consists of both thermal and nonthermal components \citep[e.g.,][]{tabatabaei17}. 
Additional deep radio data at both higher and lower frequencies are necessary to better understand this effect.
Since we measured spectral indices using the 1.4~GHz survey, which has a lower sensitivity, our measurements are biased toward steeper spectra. 
The usual assumption is that the spread in the spectral index will statistically cancel out for a large enough sample. This might not hold for the sparsely populated high-luminosity bins that at all redshifts contain a small number of galaxies (see Table~\ref{tab:lumfun_vmax}).
Almost all of the brightest galaxies ($\lum>10^{26}~\whz$) have a spectral index measurement, and 27\% of them have a very steep spectrum ($\alpha<-1.5$). 
When $\alpha=-0.7$ is assumed for all sources, the high-luminosity outliers are no longer present in the three redshift bins above $z>2$ (see Fig.~\ref{fig:lfgrid}). 
The total radio LF is also generally lower in the AGN regime, thus improving the agreement  with the individual population fits.
It is also worthwhile to mention that catastrophic failures of the photometric redshift estimates might also produce the observed outliers as photometric redshifts are used for 76\% of these bright objects.

\subsubsection{Population classification bias}

The radio excess criterion from Eq.~\ref{eq:excess_cut} was designed to select AGN with high purity; by reducing the threshold, more galaxy composites would be reclassified as AGN.
To investigate this effect, we relaxed the radio excess selection criterion from $3\sigma$ to $2\sigma$ significance. This is achieved by setting the Eq.~\ref{eq:excess_cut} parameters to $a=21.8$ and $b=0.011$. The estimated AGN contribution to the total emitted radio power in such a sample, measured from the offset from the peak of the $\log(\lum/\text{SFR}_\text{IR})$ distribution, is larger than 65\%.
We have recalculated the radio LFs of individual populations selected in this manner (shown with triangles in Fig.\ref{fig:lfgrid}). We found that the newly derived LFs are still well described with the local LFs from Eqs.~\ref{eq:lflocal_sf} and \ref{eq:lflocal_agn} evolved in luminosity. 
The AGN LF is systematically shifted to higher densities/luminosities, and the opposite is true for the SF LF. This change of the selection criterion mostly affects the faint observed end of the AGN LF and the bright end of the SF LF. These new fits fall within the $3\sigma$ shaded areas in Fig.~\ref{fig:lfgrid}, strengthening our previous results, and the LFs of the two populations classified with the $2\sigma$ selection cut are in better agreement with the fit performed on the total radio LF.
If the selection cut were relaxed even further to just $1\sigma$ radio excess (average AGN contribution larger than 40\%), our measured LFs would no longer be consistent with the local LFs. The number density of AGN at the faint end would be too high, while the bright end would remain unchanged. Therefore the $1\sigma$ radio excess cut should probably not be used when dividing the radio emission into AGN or SF dominated.

%Given that the faint luminosity end is dominated by disk galaxies and a clear cut between different populations does not exist,  we assume that the sum of two differently evolving LFs are enough to describe global trends from which we can constrain the faint radio sky. The fit is performed on the total radio LF obtained from \vmax\ and therefore all subpopulation intricacies are encoded in the shapes of the fitting functions. 

Given the dichotomy of the AGN galaxy population discussed in Sect.~\ref{sec:intro}, the AGN LF should be additionally decomposed into two separate LFs, each one evolving differently \citep[e.g.,][]{pracy16}.
Wide area observations are necessary to obtain a statistically significant sample of the brightest and the rarest objects, and the COSMOS field is not best suited for such an analysis.
Nevertheless, we attempted to fit our total radio LF with a three-population model.
The AGN below $\lum=10^{26}~\whz$ observed locally are predominantly LEGs, while HEGs would be observed at higher luminosities \citep[see][]{pracy16}. At higher redshifts the data are preferably fitted with the simple power-law used to describe the HERG population. Because of the large parameter degeneracy and insufficient number of sources, we do not report the obtained best-fit parameter values. The three-population model can explain the highest luminosity outliers, but it cannot be distinguished from possible spectral index systematics (as explained above).
Given the limitation of the data, the two population model describes our data the best. 
Our model is best understood as a tracer of the radio emission origin (star formation or AGN), even though the dominant AGN host type, as classified from optical spectra, changes with redshift \citep[see also][]{kimball11}. 
%From the radio excess perspective, our galaxies show a continuous increase of AGN contribution the further the galaxy lies from the $\lum/\text{SFR}_\text{IR}$ peak, while optically selected AGN usually exhibit radio loudness dichotomy.
Our LFs can be further used to estimate, for example, the cosmic star formation rate density or the total AGN kinetic power density.

%Here we do not consider this additional classification for the following reasons. First, we are interested in separating galaxies based on the dominant process which gives rise to the observed radio emission, because the radio power can be further scaled to estimate for example the star formation rate or the AGN kinetic power. In this scenario it is not relevant if a galaxy hosts an AGN if almost all of its radio emission is due to supernova remnants. Second, given that the VLA-COSMOS is a 2 square degrees deep survey, with our data it is not possible to constrain well the AGN LF turnover as only the power-law like faint end of the AGN population is sampled well. A wide and shallow survey is needed to fill the bright part of the LF with the rarest objects. Furthermore, the uncertainty of the radio SED shape affects mostly the bright end of the LF which presents problems in disentangling the evolution trends from possible radio spectral index systematics (see also \citetalias{novak17} and \citetalias{smolcic17c} for more detailed discussion on this), possibly explaining the outlying most luminous LF points in Fig.~\ref{fig:lfgrid}. 

The SF population would also benefit from a further decomposition into differently evolving normal/disk galaxies, starbursts/mergers and galaxies with low level AGN activity \citep[see, e.g.,][]{hopkins10}, but a classification like this is beyond the scope of this paper and would introduce too many degenerate parameters into our models.

Completeness corrections used to calculate the \vmax\ might also result in population-based biases. We assume that the completeness curve constructed in \cite{smolcic17a} equally applies to all radio sources. However, the authors found a strong dependence of the incompleteness on the intrinsic source size distribution. Specifically, if AGN were more compact and therefore easier to detect in a survey with a given surface brightness sensitivity, our AGN number densities might be overestimated. The intrinsic radio source size distributions for different classes of radio sources are still not constrained well enough to investigate this effect further and will be presented elsewhere (Bondi et al. in prep.).

\begin{table*}
\begin{center}
\caption{Best-fit PLE parameters obtained from the multivariate fitting to the total radio LF using two populations as given in Eq.~\ref{eq:lf_total}.}
\renewcommand{\arraystretch}{1.4}
\ifonecol \small \fi
\begin{tabular}{cccccc}
\hline
 Model & Description & $\alpha_L^{\text{SF}}$ & $\beta_L^{\text{SF}}$ & $\alpha_L^{\text{AGN}}$ & $\beta_L^{\text{AGN}}$ \\
\hline
% 0 &Subpopulations$^*$  & 3.16 $\pm$  0.20 & -0.32 $\pm$  0.07 & 2.88 $\pm$  0.82 & -0.84 $\pm$  0.34 \\
  0 &Subpopulations$^*$  & 3.16 $\pm$  0.04 & -0.32 $\pm$  0.02 & 2.88 $\pm$  0.17 & -0.84 $\pm$  0.07 \\
% 0 &Subpopulations$^*$  & 3.16  & -0.32 & 2.88 & -0.84 \\
 1 &PLE fit to the total LF & 2.95 $\pm$  0.04 & -0.29 $\pm$  0.02 & 2.86 $\pm$  0.16 & -0.70 $\pm$  0.06 \\
 2 &\cite{mauch07} local LFs & 2.77 $\pm$  0.04 & -0.23 $\pm$  0.02 & 3.05 $\pm$  0.15 & -0.76 $\pm$  0.06 \\
 3 &\cite{condon02} local LFs & 3.58 $\pm$  0.04 & -0.45 $\pm$  0.02 & 2.46 $\pm$  0.16 & -0.60 $\pm$  0.06 \\
 4 &\cite{parsa16} faint slope & 3.04 $\pm$  0.04 & -0.19 $\pm$  0.02 & 2.93 $\pm$  0.16 & -0.73 $\pm$  0.06 \\
\hline
\end{tabular}
\vspace{\baselineskip}\\
\scriptsize
$^\mathrm{*}$Fits on individual populations published in \citetalias{novak17} and \citetalias{smolcic17c}. We note that reported errors on the fitted parameters are corrected (reduced) here to the listed $1\sigma$ uncertainties. \\
\label{tab:evol}
\end{center}
\end{table*}

\section{Radio number counts}
\label{sec:counts}
With several LF evolution models established in previous chapters, we now aim to calculate the radio number counts down to submicrojansky flux densities.
This limit was chosen to encompass the limits of future deep radio surveys, and to enable comparison with the literature.
The number of galaxies $\Delta N$ in a given spherical shell volume $\Delta V$ and a luminosity decade $\Delta\log L$ is given by the definition of the luminosity function $\Phi$ as
\begin{equation}
\Delta N=\Phi(L, z)\,\Delta V\,\Delta \log L.
\end{equation}

%
%\subsection{Obtaining number counts from luminosity functions}
%\label{sec:counts_calc}
%
%The number of galaxies $\Delta N$ in a given spherical shell volume $\Delta V$ and a luminosity decade $\Delta\log L$ is given by the definition of the luminosity function $\Phi$ as
%\begin{equation}
%\Delta N=\Phi(L, z)\,\Delta V\,\Delta \log L.
%\end{equation}
%To obtain the differential number counts $n$ which are Euclidean normalized, i.e. weighted by flux density to the power of $5/2$ (obtained from a uniform distribution of sources in a flat universe), one must sum $\Delta N$ contributions from all redshifts and luminosities which would result in flux density range of $\Delta S$ around some $S$, formally
%\begin{equation}
%n(S)=\frac{1}{4\pi \Delta S} S^{2.5} \sum_{S}^{S+\Delta S} \Delta N (S).
%\end{equation}
%%~\text{sr}^{-1}\,\text{Jy}^{1.5}
%%S_{\text{max}}-S_{\text{min}}
%We numerically integrated all LF models described with Eq.~\ref{eq:lf_total} and listed in Table~\ref{tab:evol} within a redshift range $0<z<6$, which is constrained by our radio data. We discuss the results in the following sections.

%\subsection{Individual subpopulation fits}
%\label{sec:counts_subpop}

%We first focus on the 1.4~GHz number counts obtained from LFs reported in \citetalias{novak17} and \citetalias{smolcic17c}. These authors calculated the LFs by fitting a PLE model on individual galaxy population which were separated by using the radio-excess method (see Section~\ref{sec:lumfun_pop}). Number counts are shown in Fig.~\ref{fig:counts} along with several other works from the literature.

\begin{figure}
\centering
\includegraphics[width=\columnwidth]{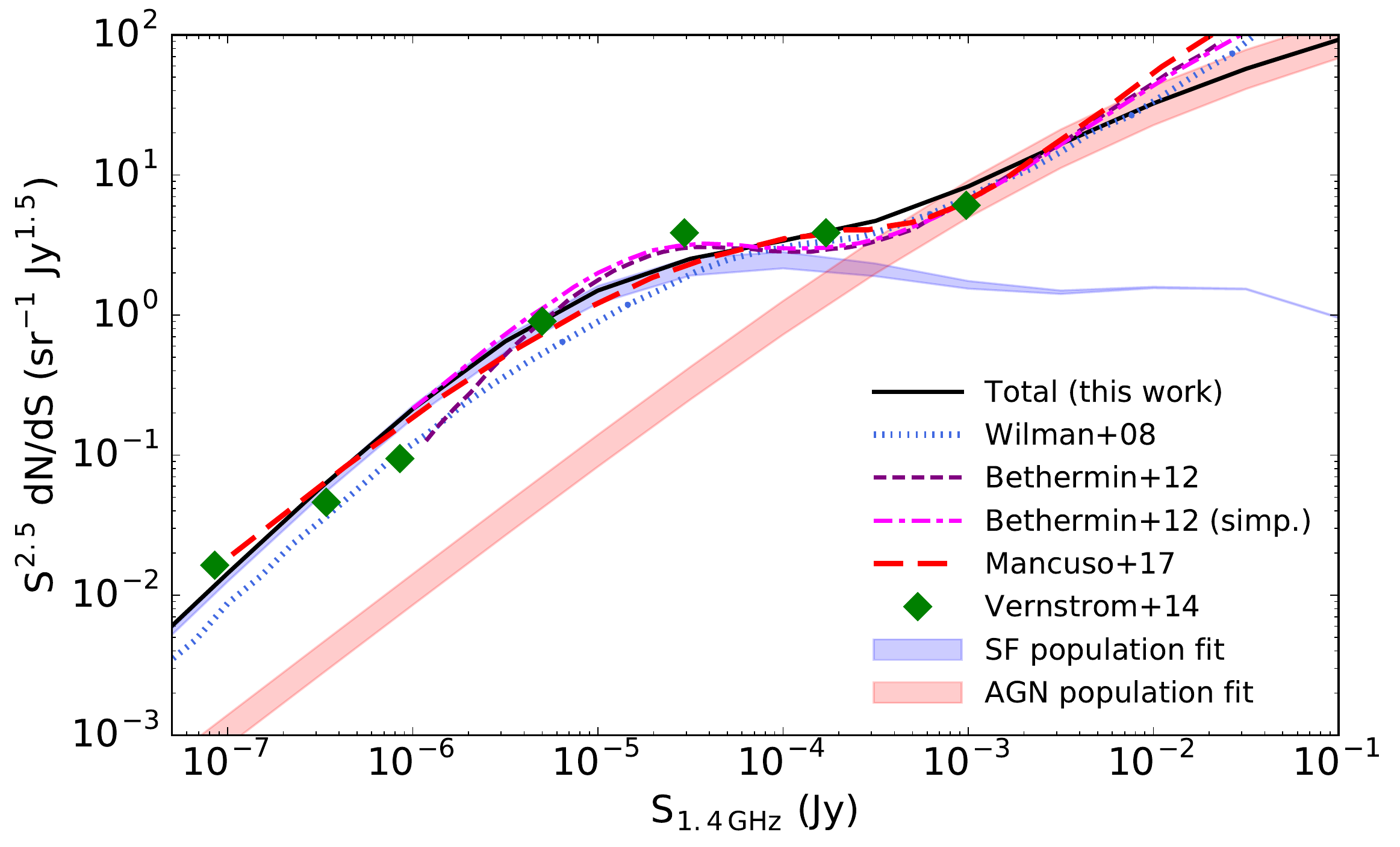}
\caption{Euclidean normalized radio source number counts drawn from LFs described in  Section~\ref{sec:lumfun_pop} compared to the literature values indicated in the legend. The shaded areas encompass the $3\sigma$ errors from the $\chi^2$ fits performed on individual populations.
}
\label{fig:counts}
\end{figure}

We compare our number counts with the results from semi-analytical models obtained by \cite{wilman08} and \cite{bethermin12} in Fig.~\ref{fig:counts}. They are all in agreement at a 1.4~GHz flux density of 100~\mujy. Our modeled number counts are systematically slightly higher at fainter flux densities (below 100~\mujy\ at 1.4~GHz) when compared to the \cite{wilman08} model, an offset probably caused by the choice of the LF and its evolution. 
Below 5~\mujy\ at 1.4~GHz our number counts are increasingly higher than the \cite{bethermin12} model, reaching a factor two difference at $\sim1~\mujy$.
On the other hand, the agreement with their simplified model is excellent in this range (see dash-dotted magenta line in Fig.~\ref{fig:counts}). 
One of the refinements that the authors implemented in their simplified model in order to obtain their final result was the dust attenuation. This in practice flattens the faint end of the IR LF and probably causes the discrepancy below 5~\mujy. 
Because of the uncertainty of dust obscuration and its dependence on redshift, the dust-unbiased aspect of radio observations provides a valuable addition to galaxy evolution studies.
Additionally, \cite{bethermin12} modeled only the X-ray selected AGN and ignored those associated with red sequence galaxies, which are easily identified in the radio band as AGN.
We also compared our number counts with the work performed by \cite{mancuso17}, who used the evolving  cosmic SFR function obtained from the FIR, ultraviolet (UV), and H$\alpha$ data to estimate radio (synchrotron and free-free emission) LFs. We find a good agreement with their work down to 0.1~\mujy.
%Below this flux density there is an increasing disagreement between the counts, however in Section~\ref{sec:counts_mcmc} we use a model with a changing faint end slope for SF galaxies, which yields higher densities of SF galaxies at submicrojansky levels.

An important result stems from the comparison of our counts with those obtained by \cite{vernstrom14}. These investigators modeled the confusion amplitude distribution $P(D)$ from deep VLA 3~GHz observations of the Lockman Hole. Their analysis uses the noise distribution of the radio map to constrain the underlying radio number counts down to $S_{3~\text{GHz}}\sim50$~nJy. A great advantage of their approach is that it is not affected by incompleteness issues from direct counting and counterpart cross-correlations, and can be used to probe the radio sky properties below the nominal sensitivity limit for source detection. 
The excellent agreement between the results from their blind method and our counts suggests that the evolving LFs published in \citetalias{novak17} and \citetalias{smolcic17c} provide plausible cosmic densities of galaxies, even though they rely on uncertain extrapolations toward the faint end. There is no need for a potential third population of radio sources (e.g., dwarf galaxies) at 0.1 -- 1~\mujy\ levels, and these LFs can reproduce the observed radio sky background.
The largest discrepancy between the \cite{vernstrom14} results and ours is present in the 10 -- 100~\mujy\ range. 
Although the formal errors on the individual fits of \cite{vernstrom14} are small, the spread can be up to 30\% depending on the area they analyze (their zones 1-3). In addition, they based their result on a field of 0.02 square degrees, which is 100 times smaller than the COSMOS field. Cosmic variance may therefore have a significant impact on their result \citep[see][]{moster11}. 
This issue has previously been identified and tested by \cite{smolcic17a}, since the flux densities above 10~\mujy\ are also above the 5$\sigma$ sensitivity limit of the VLA-COSMOS 3~GHz Large Survey (see their Fig.~18). It was shown that the sample variance obtained by splitting the COSMOS field into 100 smaller areas was enough to bring the two measurements into agreement.

Our number counts at the bright end exhibit a turnover at somewhat lower flux densities when compared to semi-analytical models (derived from abundant observations at $S_{1.4~\text{GHz}}>1~$mJy, see \citealt{dezotti10}). This is probably due to an overly simplified treatment of the brightest AGN. However, we emphasize that our sample is optimized to probe the faint radio sky and SF galaxies.
%, and extrapolations toward the faint end are discussed next.

\subsection{The faint radio sky: What will the SKA see?}

\label{sec:counts_mcmc}
\begin{figure}
\centering
\includegraphics[width=\columnwidth]{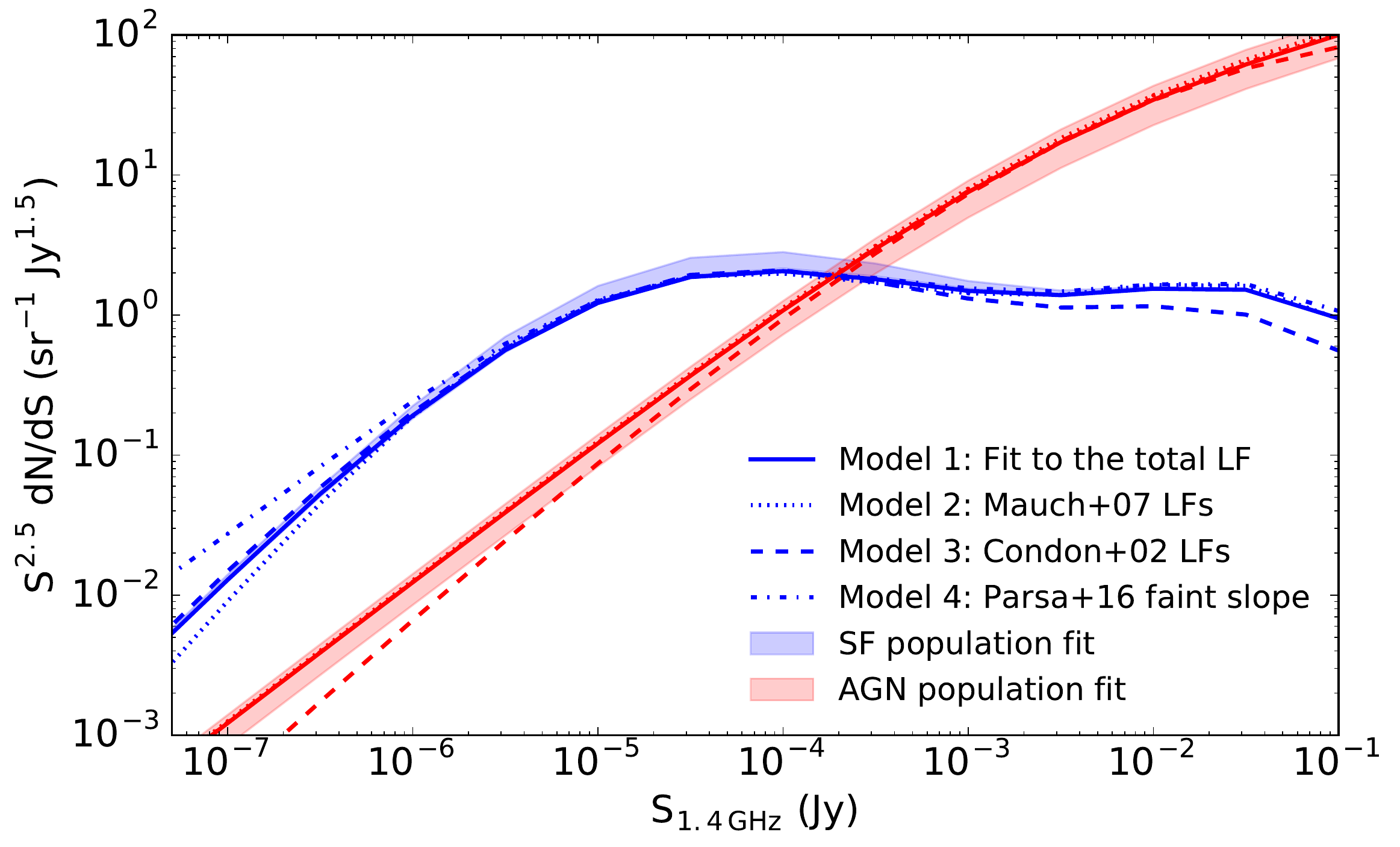}
\includegraphics[width=\columnwidth]{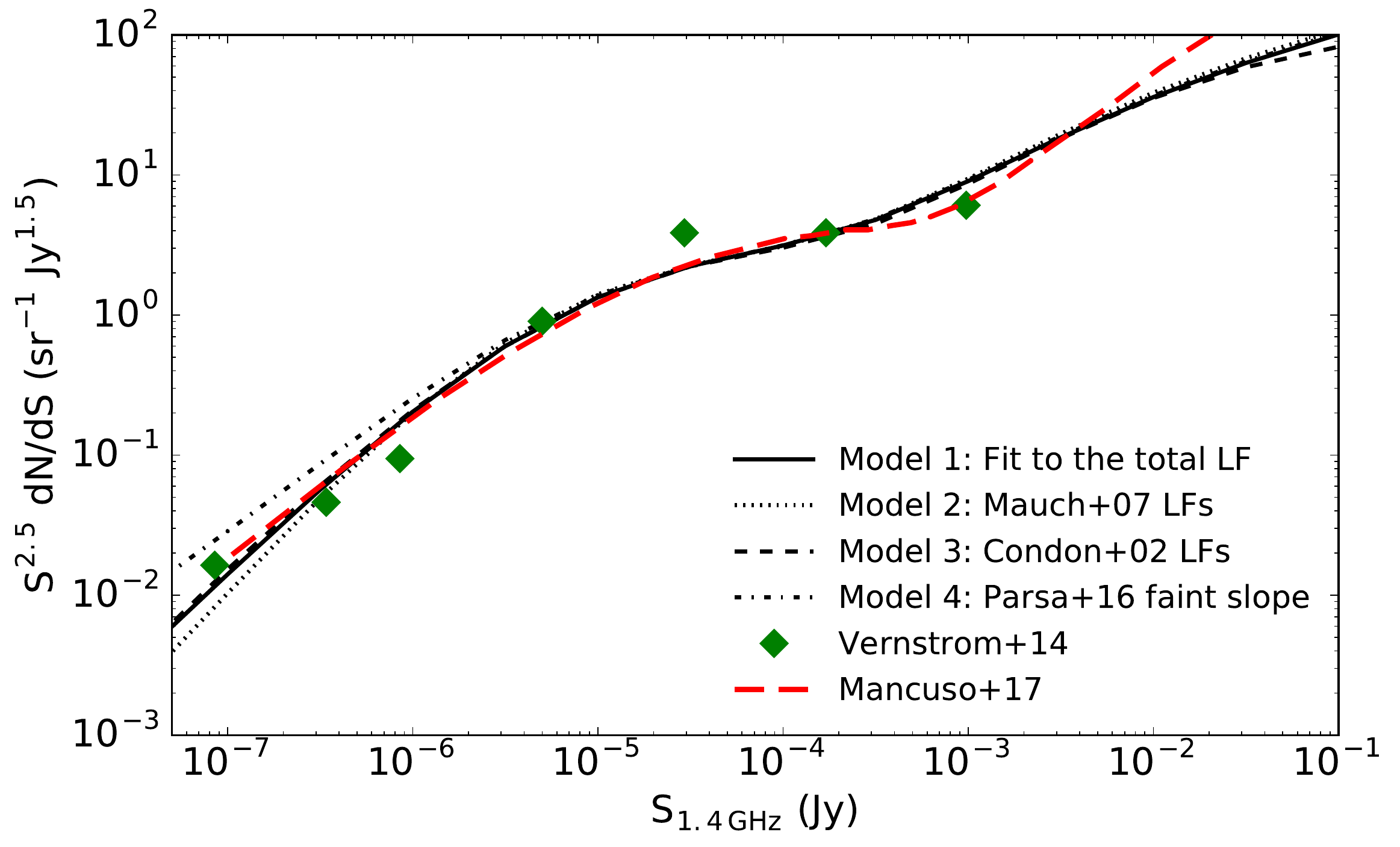}
\caption{\textit{Top panel:} Euclidean normalized radio source number counts obtained by fitting the total radio LF using different evolving analytical LFs with the MCMC algorithm. Red lines (AGN) have the same line styles as blue lines (SF), as indicated in the legend. The shaded areas are equivalent to those in Figure~\ref{fig:counts}. \textit{Bottom panel:} Total number counts for different models obtained by adding together SF and AGN contributions.
}
\label{fig:counts2}
\end{figure}

We consider the number counts obtained from the MCMC fit described in Section~\ref{sec:fit_total} (model 1) as our main result and list them in Table~\ref{tab:counts}. 
To quantify possible issues due to the faint-end slope of the LF, we tried evolving different LFs available in the literature. 
All models are shown in Fig.~\ref{fig:counts2}. Overall, they are consistent among themselves, although with some expected variations below 1~\mujy. 

%Due to the almost flat faint power-law slope of the SF LF reported in \cite{mauch07}, and a steep one for the AGN, this model predicts the fewest SF galaxies at nano-Jy flux densities. This is not the case with \cite{condon02} local LFs where the slopes for both populations are similar leading to a roughly constant ratio of SF to AGN number counts below 1~\mujy.

By comparing the radio and the UV LFs, \citetalias{novak17} noted that the evolved local radio LF might have a  faint-end slope (below $\lum<10^{22}$) that is too flat when compared to UV LFs at high redshift ($z>3$). To obtain some insight into this effect, we have constructed a fourth model that uses a redshift-dependent faint-end slope, derived by \cite{parsa16}, to describe UV LFs. This model is the same as our first model, with the exception that the faint slope parameter for SF galaxies ($\alpha$ in Eq.~\ref{eq:lflocal_sf}) now changes linearly with redshift as $\alpha=0.106z+1.187$. As expected, the result of this LF choice is an increased density of faint SF galaxies, higher than either the \cite{mancuso17} or \cite{vernstrom14} estimates.
%, which is then reflected in an order of magnitude larger number counts at 1 nJy than the values obtained from the first model. In this case the SF population dominates the radio number counts down to nJy levels, unless the AGN faint slope also steepens considerably.

\begin{table*}
\begin{center}
\caption{Euclidean normalized differential radio number counts as a function of flux density at 1.4~GHz.}
\renewcommand{\arraystretch}{1.4}

%\small
\begin{tabular}{cccc}
\hline
$\log(S_\text{1.4~GHz})$ & $\log(n^{\text{SF}})$ & $\log(n^{\text{AGN}})$ & $\log(n^{\text{Total}})$ \\
 $[$Jy$]$&$[$sr$^{-1}$\,Jy$^{1.5}]$&$[$sr$^{-1}$\,Jy$^{1.5}]$&$[$sr$^{-1}$\,Jy$^{1.5}]$\\
\hline
-7.5 & -2.52 & -3.42 & -2.47 \\
-7.0 & -1.89 & -2.91 & -1.85 \\
-6.5 & -1.28 & -2.41 & -1.25 \\
-6.0 & -0.717 & -1.91 & -0.690 \\
-5.5 & -0.250 & -1.41 & -0.221 \\
-5.0 & 0.0865 & -0.917 & 0.128 \\
-4.5 & 0.271 & -0.434 & 0.350 \\
-4.0 & 0.314 & 0.0330 & 0.497 \\
-3.5 & 0.256 & 0.474 & 0.680 \\
-3.0 & 0.173 & 0.878 & 0.956 \\
-2.5 & 0.142 & 1.23 & 1.27 \\
-2.0 & 0.188 & 1.54 & 1.56 \\
-1.5 & 0.180 & 1.79 & 1.80 \\
-1.0 & -0.0234 & 2.00 & 2.00 \\

\hline
\end{tabular}
\label{tab:counts}
\end{center}
\end{table*}

\begin{figure}
\centering
\includegraphics[width=\columnwidth]{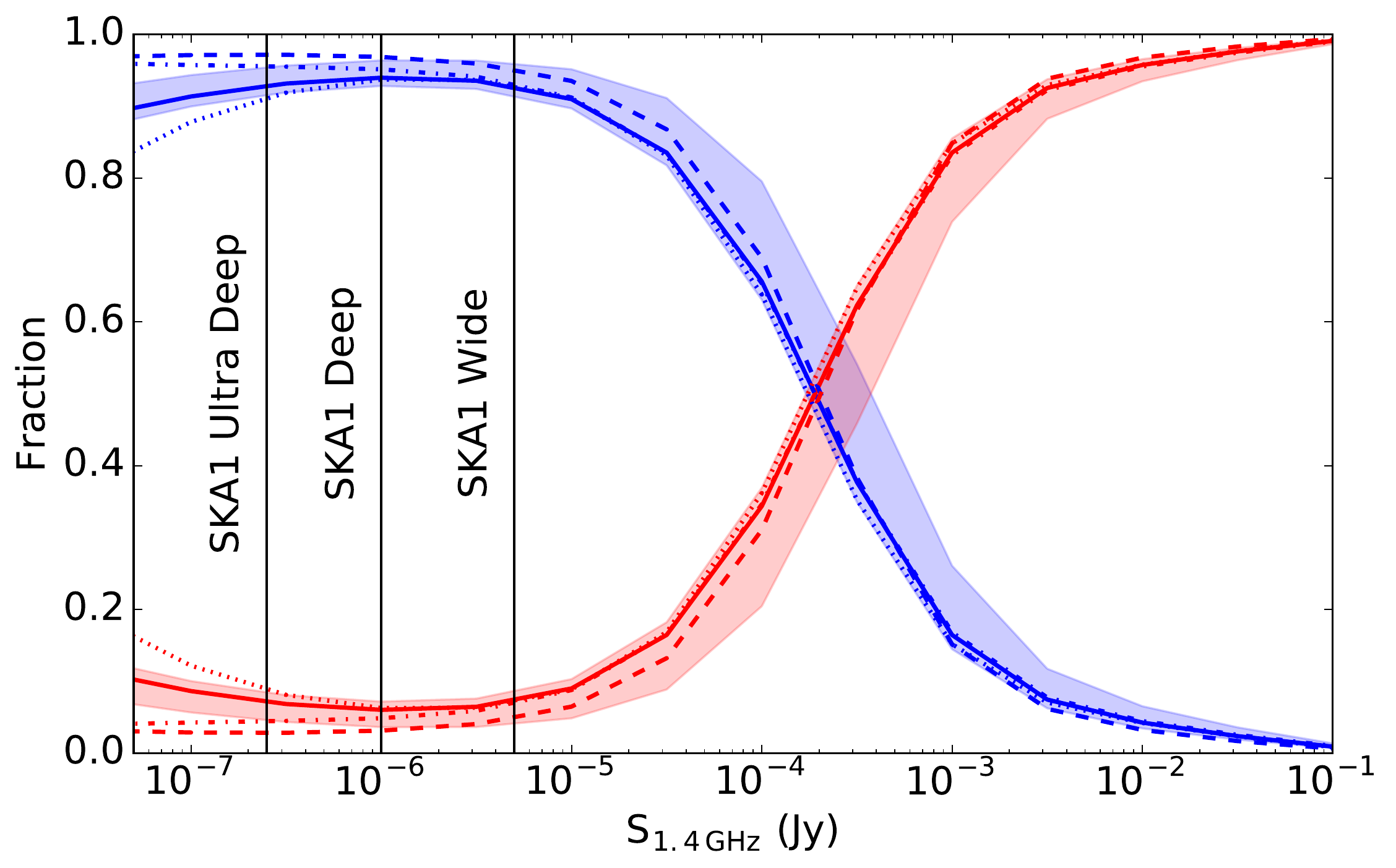}
\caption{Percentages of AGN (red) and SF galaxies (blue) as a function of 1.4~GHz flux density. Colors and line styles are the same as in Figure~\ref{fig:counts2}.}
\label{fig:fraction}
\end{figure}

In Fig.~\ref{fig:fraction} we show the fraction of each population as a function of flux density to better illustrate the effect of how relative abundances of SF galaxies and AGN change with telescope sensitivity. Several conclusions can be drawn from this plot. Above 1~mJy at 1.4~GHz, the majority of radio emission is due to AGN, as is well known from past studies. 
At around 200~\mujy\ at 1.4~GHz, both populations contribute
equally to the observed counts, and at fainter flux densities, SF galaxies become the dominant population.
%All LF models agree down to 1~\mujy, however at 1-100~nJy they diverge due to different faint end slopes of the LFs.
%This divergence also causes the AGN percentage to rise in some models at faintest flux densities. The model based solely on \cite{mauch07} LFs predicts that AGN and SF galaxies will again be equally abundant at 1~nJy, however this is a product of extrapolating the LFs with significantly different faint end slopes (the SF LF slope is flat and the AGN LF slope is steep). Observations at UV wavelengths do not support such a flat faint end slope of the SF LF \citep[e.g.][]{bouwens15}.
The most important result is that between 0.1 and 10~\mujy, where the $5\sigma$ sensitivity limits for future SKA surveys at 1.4~GHz are located \citep[see][]{prandoni15}, the relative fraction is roughly constant with around 90-95\% of the radio emission originating from star formation. The implication of this result is that a simple flux density cut at $S_{1.4~\text{GHz}}<10~\mujy$ will select samples with less than 10\% contamination by AGN radio emission, thus providing constraints on the radio emission origin.

%Applying such a cut on a radio selected sample with flux densities from 1~mJy down to 1~\mujy\ (0.1~\mujy) would yield 88\% (97\%) completeness of the SF galaxy sample, based on the integrated source number counts.
%with the remaining galaxies having $S_{1.4~\text{GHz}}>10~\mujy$, based on the integrated source number counts. 
%The above statement assumes that the radio telescope can observe the total integrated emission because the AGN component is compact and star-formation is usually smeared across the disk. 
%Whether the assumption is valid depends on the beam size and the galaxy surface brightness.

\begin{figure}
\centering

\includegraphics[width=\columnwidth]{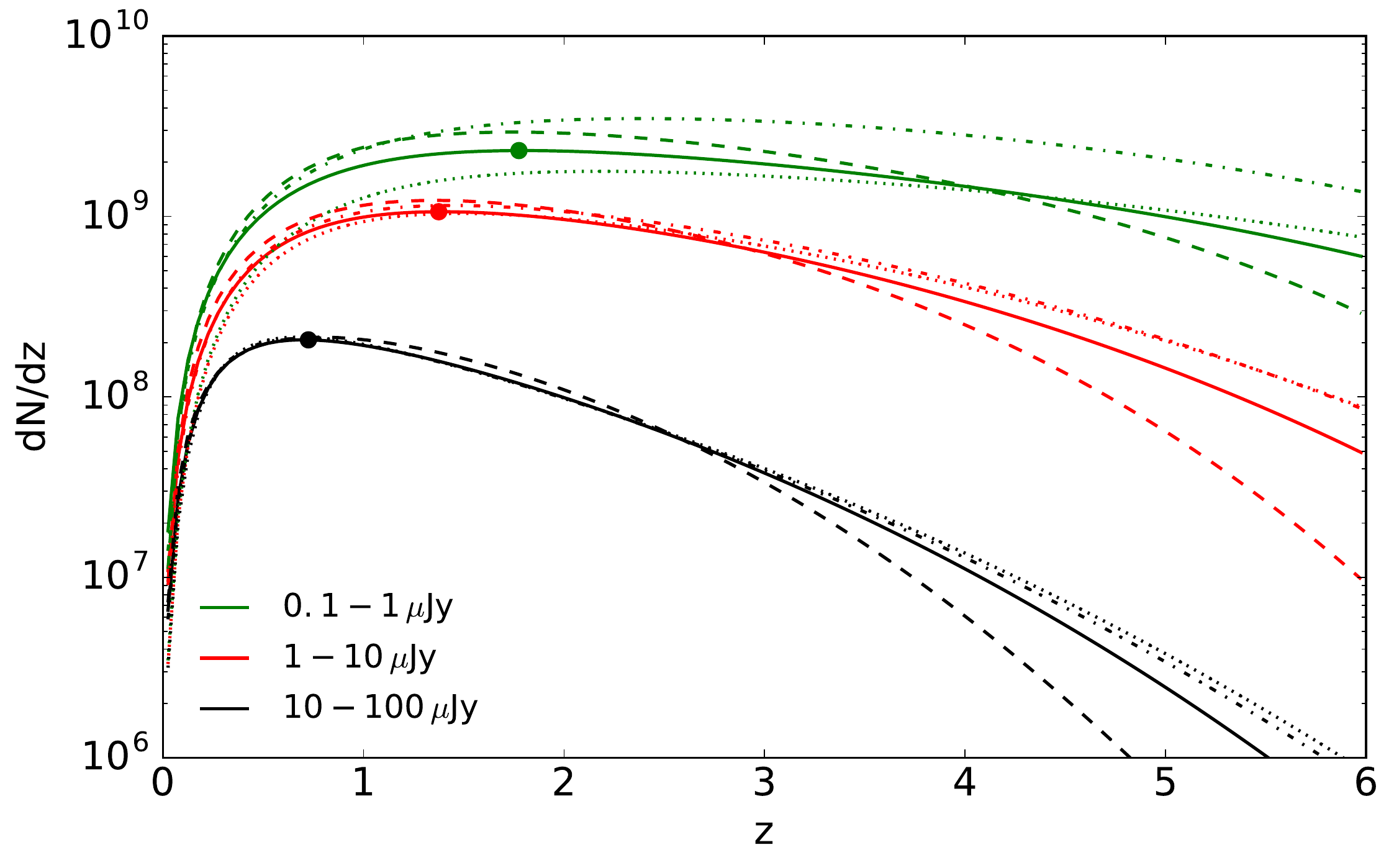}

\caption{Redshift distributions (differential numbers across the full sky, i.e., 4$\pi$ solid angle) of SF galaxies in different flux density decades based on evolving radio LFs at 1.4~GHz. Filled circles show the peaks of model~1. Different line styles describe models 1 -- 4 as in Figure~\ref{fig:counts2}.}
\label{fig:redshift_distribution}
\end{figure}

We further analyzed our different models of evolving radio LFs and derived the redshift distribution of SF galaxies in different flux density ranges as presented in Fig.~\ref{fig:redshift_distribution}. Different models can yield similar number counts, but with different intrinsic redshift distributions. The VLA-COSMOS 3~GHz Large Project survey has a median $5\sigma$ sensitivity of 11.5~\mujybeam\ and around 90\% of all radio sources with optical/NIR counterparts fall inside the 
1.4~GHz flux density range of 10 -- 100~\mujy. As can be seen from the figure, a large portion of our observed galaxies is distributed around a redshift of $z\sim1$.
% All our models predict that,  with an increase in sensitivity, the bulk of all observed galaxies will shift toward a redshift of $z\sim2$.
Our models predict that only by probing the submicrojansky population (0.1 -- 1~\mujy) a significant shift of the source redshift distribution (peak at $z\sim 2$) can be obtained.

%Models 1 and 2 have a flatter faint end LF slope for SF galaxies and suggest that, even with an increase of several decades in sensitivity, the peak number of all observed galaxies will remain around $z\sim2$. On the other hand, if there is a significant trend which favors much larger densities of the faintest galaxies at higher redshifts, such as the model 4, then the peak of the galaxy number distribution also shifts toward higher redshifts with each increase in sensitivity. Regardless of the peak position, at sensitivities below 10~\mujy\, all distributions have extended tails toward higher redshifts and significantly higher overall  numbers of galaxies. To give an example, future deep radio observations will yield as many redshift $z\sim4$ SF galaxies as current surveys find at $z\sim1$.

\subsection{Future surveys across radio bands}

The Euclidean normalized number counts $n$ at 1.4~GHz reported in Table~\ref{tab:counts} can be converted into any other radio frequency $\nu$ under an assumption of a simple power-law radio SED with a spectral index $\alpha$ (usually $\alpha=-0.7$) taking $S_\nu=S_\text{1.4~GHz}/f$ and $n_\nu=n_\text{1.4~GHz}/f^{1.5}$, where $f=(1.4 / \nu~\text{[GHz]})^\alpha$ is the conversion factor, $n_\nu$ ($S_\nu$) are the number counts (flux densities) at the frequency $\nu$ one converts to. While this conversion is probably correct on average for most situations, it might be biased if the average radio SED of a galaxy population deviates from a simple power law. For example, thermal free-free emission might provide a significant contribution to the total radio emission in SF galaxies at frequencies above 10~GHz. Below 150~MHz, synchrotron self-absorption might also affect the radio source counts. Additionally, radiative losses can impact the high-frequency counts of AGN. As deeper surveys across a wide radio bandwidth become available in the future, a comparison with this simple approach may be helpful to  better quantify biases of radio SED assumptions.

\begin{figure}
\centering
\includegraphics[width=\columnwidth]{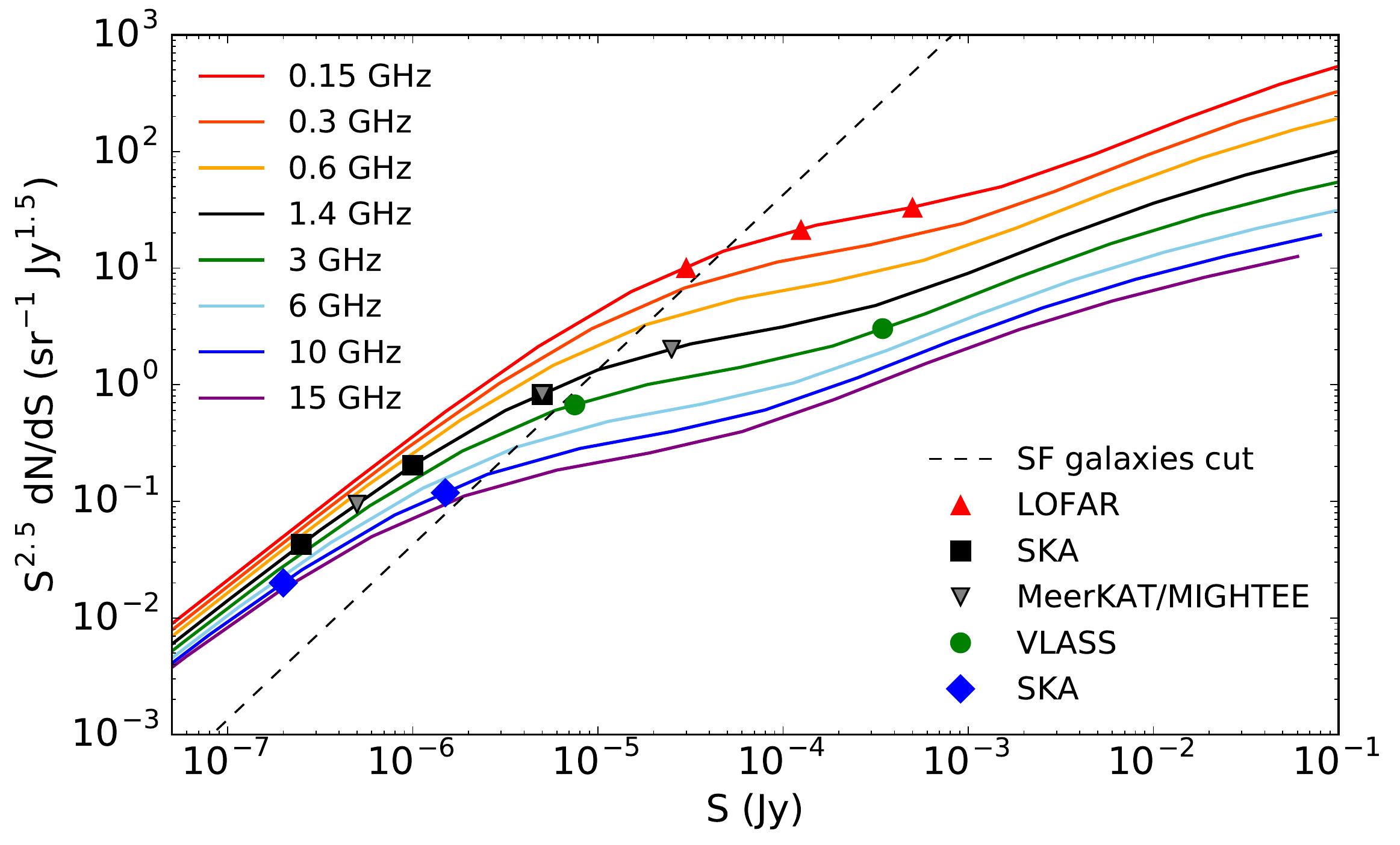}
\caption{Euclidean normalized radio number counts at different frequencies scaled from 1.4~GHz with a simple power-law radio SED (see text for details). The dashed line marks the $S_\text{1.4~GHz}=10~\mujy$ cut, below which SF galaxies highly dominate the number counts. The filled symbols represent the $5\sigma$ sensitivity limits for proposed multi-tiered (ultra-deep, deep, and wide) future radio surveys.}
\label{fig:counts3}
\end{figure}

We plot the number counts from our model 1 scaled to different frequencies in Fig.~\ref{fig:counts3}, where we also show the $5\sigma$ sensitivities of several planned radio continuum surveys \citep[see also][]{norris13, prandoni15}. All of these surveys have multiple tiers stemming from the "wedding cake" observing approach: from wide and shallow observations spanning thousands of square degrees, to pencil-beam-deep and ultra-deep observations. 
In the near future, observations such as the LOFAR\footnote{The Low-Frequency Array} survey at 150~MHz (see \citealt{morganti10} and also \citealt{shimwell17}) or the VLASS\footnote{The Very Large Array Sky Survey \url{https://science.nrao.edu/science/surveys/vlass}} at 3~GHz will be able to provide a solely radio-based selection of SF galaxies only at the sensitivity limits of their deep tier observations.

%The SKA, and its precursors, will provide an additional method to classify galaxies as star-forming across large areas of the sky.

%The flux density cut for selecting SF galaxies can be increased at the cost of sample purity, for example, a $S_\text{1.4~GHz}<45~\mujy$ cut will yield up to 20\% contamination by AGN dominated galaxies. 
%The strength of this approach is that multi-wavelength photometry and dust-corrections are no longer necessary to perform AGN versus SF galaxy classification in the color-color space or with the SED fitting. 

%in a similar manner that the Very Large Baseline Array (VLBA) can confirm AGN, only across a much larger area of the sky.

\section{Summary}
\label{sec:summary}
We used the VLA-COSMOS 3~GHz Large Project radio data \citep{smolcic17a} to construct the total radio luminosity functions extending out to redshift $z<5.7$ from a sample of 7826 radio sources within an area of 1.77~deg$^2$ with robust optical/NIR counterparts. We fitted the total radio LF with pure luminosity evolution models that depend on redshift. They rely on the sum of differently evolving SF and AGN LFs. We tested LFs of star-forming and AGN galaxies published in \cite{novak17} and \cite{smolcic17c} and found them to be consistent with the Markov chain Monte Carlo fits performed in this paper. 
We have tested our radio excess selection criteria against the total radio LF, and have found that both $3\sigma$ and $2\sigma$ cuts are consistent with our previously published LFs.
The total radio LF is in better agreement with the $2\sigma$ cut.
However, it is difficult to distinguish  selection effects from potential spectral index systematics.
We calculated radio number counts from our evolution models down to submicrojansky levels. We found that our LFs can well reproduce the number counts obtained from a blind probability of deflection $P(D)$  analysis \citep{vernstrom14}, implying that our extrapolations toward the faint end of the LFs are plausible even though they are not constrained directly by the data. Finally, we showed that planned surveys with the SKA will almost exclusively probe the star-forming galaxy population with a peak in the redshift distribution at $z\sim2$. Our results suggest that the radio sky between 0.1 and 10~\mujy\ at 1.4~GHz is dominated by star formation processes with a maximum of  $\sim$10\% contamination from the radio AGN population. Future radio surveys will be able to provide an additional simple diagnostic for classification of galaxies across large areas based solely on radio flux densities.
%, thus enabling a simple cut in flux density to select a star-forming galaxy, instead of the usual multi-wavelength color-color  separation approach. \mn{Phrasing?}

\begin{acknowledgements}
We thank the anonymous referee for helpful  suggestions that improved the quality of the paper.
This research was funded by the European Unions Seventh Frame-work program under grant agreement 337595 (ERC Starting Grant, \textit{CoSMass}).

\end{acknowledgements}

\bibliographystyle{aa}
\bibliography{bibtex}

\end{document}